\documentclass[letter]{aa} % for the letters 
\usepackage{graphicx}
\usepackage{txfonts}
\usepackage{color}
\begin{document}
   \title{CAL 87 -- an evolved wind-driven supersoft X-ray binary
	\thanks{Based on observations made at the SOAR Telescope.}
	}

   \author{A. S. Oliveira \inst{1,2}
  \and
  J. E. Steiner \inst{3}
          }

   \institute{IP\&D, Universidade do Vale do Para\'{\i}ba, Av. Shishima Hifumi, 2911, CEP 12244-000, S\~ao Jos\'e dos Campos, SP, Brasil \\
	     \email{alexandre@univap.br}
	     \and
	      SOAR Telescope, Casilla 603, La Serena, Chile
	     \and
	     Instituto de Astronomia, Geof\'{\i}sica e Ci\^encias Atmosf\'ericas, Universidade de S\~ao Paulo, 05508-900,
    S\~ao Paulo, SP, Brasil\\
             \email{steiner@astro.iag.usp.br}
             }

   \date{Received / Accepted}

% \abstract{}{}{}{}{} 
% 5 {} token are mandatory
 
  \abstract
  % context heading (optional)
  % {} leave it empty if necessary  
   {Compact binary supersoft X-ray sources (CBSS) are explained as being associated with hydrostatic nuclear burning on the surface of a white dwarf with high accretion rate. This high mass transfer rate has been suggested to be caused by dynamical instability, expected when the donor star is more massive than the accreting object. When the orbital period is smaller than $\sim$6 hours, this mechanism does not work and the CBSS with such periods are believed to be fed by a distinct mechanism: the wind-driven accretion. Such a mechanism has been proposed to explain the properties of objects like \object{SMC 13}, \object{T Pyx} and \object{V617 Sgr}. One observational property that offers a critical test for discriminating between the above two possibilities is the orbital period change.}
  % aims heading (mandatory)
   {As systems with wind-driven accretion evolve with increasing periods, some of them may reach quite long orbital periods. The above critical test may, therefore, also be applied to orbital periods longer than 6 hours. \object{CAL 87} is an eclipsing system in the LMC with an orbital period of 10.6 hours that could provide the opportunity for testing the hypothesis of the system being powered by wind-driven accretion.}
  % methods heading (mandatory)
   {We obtained eclipse timings for this system and show that its orbital period increases with a rate of  $P/\dot{P} = +7.2(\pm1.3) \times 10^{6}$ years.}
  % results heading (mandatory)
   {Contrary to the common belief, we conclude that CAL 87 is the first confirmed case of a wind-driven CBSS with an orbital period longer than 6 hours. The system is probably an evolved object that had an initial secondary mass of $M_{2i}=0.63 M_{\sun}$ but is currently reduced to about $M_{2}=0.34 M_{\sun}$.}
  % conclusions heading (optional), leave it empty if necessary 
   {We discuss evidence that other CBSS, like \object{CAL 83} and V Sge stars, like \object{WX Cen}, are probably also wind-driven systems. This may in fact be the rule, and systems with inverted mass ratio, the exception.}

   \authorrunning{A. S. Oliveira \& J. E. Steiner}
   \titlerunning{CAL 87 -- A wind-driven supersoft X-ray binary}
   \maketitle
%
%________________________________________________________________

\section{Introduction}

Compact binary supersoft X-ray sources (CBSS) are a group of objects, first discovered in the Magellanic Clouds, with unusual
properties. Their large ($\sim$Eddington) luminosity is mostly radiated in the supersoft X-ray spectral range of 20 to 80 $eV$.
Van den Heuvel et al. (\cite{heuvel}) have shown convincingly that this emission is originated from hydrostatic nuclear burning on the
surface of a white dwarf. 
In order for this to happen, a high accretion rate ($\sim10^{-7} M_{\sun}~yr^{-1}$) is required, such as may be realized in systems with inverted mass ratios, in the sense that the mass donor is the more massive of the binary components. Mass transfer is then dynamically unstable and occurs on the Kelvin-Helmholtz time-scale, with the orbital period decreasing with time.
For mass donor stars with masses in the range 1.3 to 2.5 $M_{\sun}$, this configuration will provide the required accretion rate of
$\sim10^{-7} M_{\sun}~yr^{-1}$. This framework has been adopted in the literature since it was proposed (van den Heuvel et al. \cite{heuvel}; see also Kahabka \& van den Heuvel \cite{kah}, for a review and references).

The short period CBSS \object{1E0035.4-7230}, also known as SMC 13 (Schmidtke et al. \cite{schmidtke2}), does not fit into this classical picture of supersoft X-ray sources as being driven by dynamically unstable mass transfer. This binary has an orbital period
of 4.1 hours and the mass donor star is certainly less massive than the white dwarf. In this case 
we have a mass ratio that is
normal for cataclysmic variables, and the dynamical instability does not occur. How can we explain the required high accretion rate? Van
Teeseling \& King (\cite{tees}) have proposed that high mass transfer rate could be driven from a strongly irradiated low-mass donor star.
Such an explanation was proposed for SMC 13 (van Teeseling et al. \cite{tees}) and for T Pyx (Knigge et al. \cite{knigge}), and could also
be applied to \object{GQ Mus} in its long post-nova phase  (Diaz \& Steiner  \cite{diaz89}; Diaz \& Steiner
\cite{diaz94}). A critical test that could be applied to the two groups of stars is to measure the orbital period change: the (high
mass) dynamical instability mass transfer (DIMT) model predicts an orbital period decrease while the (low mass) wind-driven mass transfer
(WDMT) model predicts a period increase.

This critical test was applied to the V Sagittae star V617 Sgr (Steiner et al. \cite{stei06}). V Sge stars are considered the galactic
counterpart of the CBSS (Steiner \& Diaz \cite{stei98}). All three well known stars of this class have eclipses, and this provides a good
opportunity for measuring variations of the orbital period. \object{V Sge} has an orbital period of 12.3 hours and a decreasing orbital period (Patterson et al. \cite{patter}); this is consistent with the DIMT scenario. \object{V617 Sgr} has an orbital period of 
4.98 hours (Steiner et al. \cite{stei99}) and, therefore, is below the critical period of 6 hours for the DIMT model 
(Deutschmann \cite{deutsch}; King et al. \cite{king2}). In fact, an increasing orbital period has recently been reported (Steiner et al. \cite{stei06}), confirming the WDMT scenario. In
such a case, the orbital period increases with time, as the secondary ``evaporates'', and may reach values as high as 30 hours 
(van Teeseling \& King \cite{tees}).  

One may, therefore, ask whether the other known CBSS stars with orbital periods longer than 6 hours are in fact
DIMT cases or advanced stages of WDMT? Of course, each case would have to be tested in order to verify this dichotomy. Perhaps the
most promising case to be studied in order to answer this question is \object{CAL 87}, a well known binary in the LMC 
(Cowley et al. \cite{cow}, Hutchings et al. \cite{hutchings}). Accurate timings have been measured by Cowley et al. (\cite{cow}) and, 7.1 years later, by Alcock et al. (\cite{alco}), allowing a precise determination of the orbital period. As 9.8 years have elapsed since the last timing, with an additional determination one could make a second  accurate measurement of the period, allowing assessment of any change.
In the present paper we report timings for the eclipses of CAL 87 and show that the star has an orbital period that increases with time.

\section{Observations and data analysis}

CAL 87 was observed with the SOAR Optical Imager (SOI) on six nights in October and November 2005, during the Brazilian time of the early science phase of the SOAR Telescope.
The data consist of time series of images obtained with the mosaic of two EEV $2048\times4096$ thinned and back-illuminated CCDs, through the Johnson V filter, and with individual exposure times of 60 seconds. Each series spanned about 1 hour and was centered on the expected time of the eclipses. The data reduction -- bias and flatfield corrections -- was performed with IRAF~\footnote{IRAF is distributed by the National Optical Astronomy Observatories,
which are operated by the Association of Universities for Research in Astronomy, Inc., under cooperative agreement
with the National Science Foundation.} standard routines. Differential photometry of the images was executed with the aid of the DAOPHOT II package.

From the six observed light curves, two were discarded because of the presence of flares that disturbed the measurements of the eclipse timings. The remaining four eclipse timings were measured individually, as well as combined in a single set folded in phase with the ephemeris from Alcock et al. (\cite{alco}). The average timing obtained from our data is presented in Table\ref{timings} in addition to the eclipse timings found in the literature.

\begin{table}[!h]
\caption{Eclipse timings of CAL 87.}
\label{timings}
\begin{flushleft}
\begin{tabular}{llll}
\hline
\hline\noalign{\smallskip\smallskip}
            \noalign{\smallskip}
  Obs. minimum   &E         & Reference\\
  (HJD)          & (cycles) &          \\
\hline
\noalign{\smallskip}
  2~447~506.8021(2)	   	& 0 	& Cowley et al. \cite{cow}      \\
  2~450~111.5144(3)		& 5884	& Alcock et al. \cite{alco}     \\
  2~453~680.8244(3)		& 13947 & this work			  \\
\noalign{\smallskip}
\hline
\end{tabular}
\end{flushleft}
\smallskip\noindent
\end{table}

The ephemeris of Alcock et al. (\cite{alco}) provides a period of $P_{1}= 0.442\,677\,14(6)~d$; this period was found from observations  spanning a total
interval of 5884 cycles, with the mean cycle of $E_{1} = 2942$. From our timings we can derive a period of $P_{2} = 0.442\,677\,66(5)~d$
 with respect
to cycle 5884; this corresponds to the mean cycle $E_{2} = 9915$. The difference between $P_{2}$ and $P_{1}$ is
9 to 10 times the uncertainty in the determination of those periods.  From the difference between $P_{2}$
and $P_{1}$ and the elapsed time $P.(E_2-E_1)$ we obtain 
$\dot{P} = +1.7(\pm 0.3) \times 10^{-10}$, the uncertainty of which was estimated with the usual error propagation method from the errors of $P_{1}$ and $P_{2}$. As expected, the new timing detemination allowed the measurement of the orbital period change: the
period increases with a time-scale of $P/\dot{P} = +7.2(\pm1.3) \times 10^{6}$ years.

\begin{figure}%[p]
      \resizebox{\hsize}{!}{\includegraphics[clip]{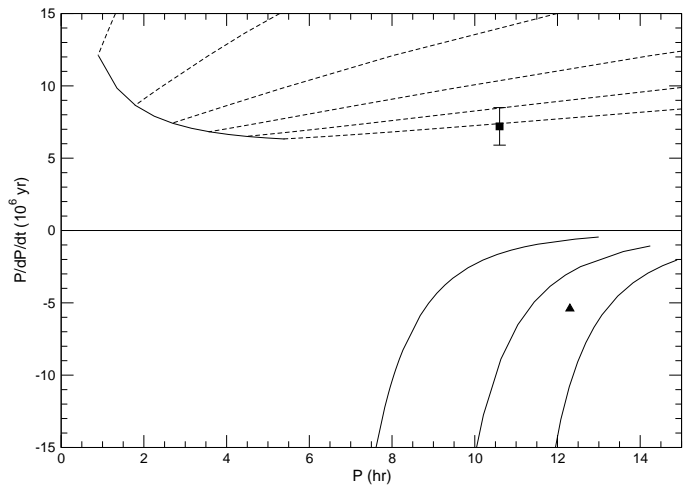}}
           \caption{$P/\dot{P}$ as a function of the orbital period. \textit{Top -- WDMT systems:} the
     solid curve represents the expected loci for systems at the beginning of the 
    wind-driven evolution. The dashed lines (from left to right) are the evolutionary tracks 
    for systems with initial secondary masses between 0.1 and 0.6 $M_{\sun}$, in steps of 0.1 $M_{\sun}$.
     The square shows the WDMT binary CAL 87 -- see Table~\ref{measured}. 
     \textit{Bottom -- DIMT systems:} the solid curves (from left to right) 
     represent the expected loci for systems with primary star masses of 0.55, 1.0 and 
     1.4 $M_{\sun}$. The triangle shows the DIMT binary V Sge.}
     \label{fig1}
  \end{figure}

\section{Assisted stellar suicide -- a terminal case?}

The first conclusion is that, contrary to the canonical view that considers CBSS as having massive (1--2 $M_{\sun}$) 
secondary stars, CAL 87 clearly
has a mass ratio which is typical for cataclysmic variables, i.e., $M_2<M_1$. This has been anticipated by
Cowley et al. (\cite{cow}, \cite{cow98}) who found, given the small amplitude of the radial velocity of emission lines, that $M_2 = 0.4 M_{\sun}$ and $M_1 > 4 M_{\sun}$. For this reason those authors explored the idea that such a massive primary could be a black hole. We now know that the mass accreting star is probably a white dwarf:
its mass, derived from fitting the luminosity-temperature diagram with theoretical calculations, is $M_1=1.35 M_{\sun}$ (Starrfield et al. \cite{starr}).
The observed supersoft X-ray luminosity, $L_X = 4 \times 10^{36}$ erg/s (Starrfield et al. \cite{starr}), implies an accretion rate of $\dot{M}_{acc}= 10^{-8} M_{\sun}~yr^{-1}$. This is additional information which can be used to solve the following equations (van Teeseling \& King \cite{tees}; see also eqs. 3, 4 and 5 in Steiner et al. \cite{stei06}):

\begin{equation}
\label{eq:um}
\frac{gq}{1+g} = \frac{\dot{M}_{acc}}{M_1} \frac{P}{\dot{P}}
\end{equation}

and

\begin{equation}
\label{eq:dois}
g^{2}\left(\frac{q^{5/2}}{1+q}\right)^{2/3} = \frac{\dot{M}_{acc}}{1.2 \times 10^{-6}~M_1} ~~~
\end{equation}

where

\begin{equation}           
\label{eq:tres}
g=\frac{3\beta_2 - q-2}{(1+q)(2-3q)}.   
\end{equation}

Here $\beta_2$ is the ratio of the specific angular momentum of the wind to that of the secondary star. 
We have also assumed that  $M_2 < 0.6$ (see van Teeseling \& King \cite{tees} for a detailed discussion).

The right-hand side of equations (\ref{eq:um}) and (\ref{eq:dois}) are measured quantities, while from the left-hand side two independent parameters can be determined: $q = 0.25$ and $\beta_2 = 0.89$. This implies that the mass of the secondary is $M_2 = 0.34 M_{\sun}$.

One should explore the range of parameters that could be accommodated with the observations. The observed supersoft X-ray luminosity means that $\dot{M}_{acc}$ cannot be smaller than $10^{-8} M_{\sun}~yr^{-1}$; therefore the minimum values for $\beta_2$ and $q$ are 0.89 and 0.25, respectively. A larger value for $\dot{M}_{acc}$ could be possible if part of the X-rays are absorbed, for instance, due to the high inclination. However, equation (\ref{eq:tres}) dictates that $q<2/3$. For this reason $\dot{M}_{acc} < 1.9 \times 10^{-8} M_{\sun}~yr^{-1}$, but $M_{2}$ could be as high as $M_{2}=0.9~M_{\sun}$. For all the range of possible parameters, we always obtain $\beta_2=0.89$.
	
Is this value for $\beta_2$ satisfactory? We adopt the idea that the wind from the secondary star must be blown from its heated face. We will make the approximation that the wind's average specific angular momentum is that of the point which is at $R_{2}/3$ from the center of the secondary star, and $\sim2/3$ of $R_{2}$ from the inner Lagrangian point. This is approximately 

\begin{equation} 
\label{eq:quatro}    
\beta_2 \sim 1 - (0.13 + 0.07~\mathrm{log}~q)(1+q)
\end{equation}

Here we have assumed that the radius of the secondary is given by the well known Paczynski's relation $R_{L2}/a = 0.38 + 0.2~\mathrm{log}~q$ (Paczynski \cite{pacz}), where $a$ is the separation between the two stars. For a mass ratio  of $q=0.25$, we obtain $\beta_2 = 0.89$, in agreement with that determined above. For higher values of the mass ratio, $\beta_2$ could be as low as $\beta_2=0.80$ (for $q=2/3$).
	The numbers derived above are, therefore, consistent; they also imply that,
for a binary with an orbital period of 10.6 hours, the secondary is quite evolved. From van Teeseling \& King 
(\cite{tees}) we have

\begin{equation}
\label{eq:quatro}
P_i \simeq 3 \sqrt{P_{orb}M_2}  ~\textrm{hr}
\end{equation}

and

\begin{equation}
\label{eq:cinco}
M_{2i} \simeq P_i/9
\end{equation}

\noindent where $P_i$ and $M_{2i}$ are the orbital period (in hours) and the secondary mass 
(in solar masses) at the beginning of the wind-driven evolution, respectively.

This means that the initial orbital period was $P_i = 5.7$ hr and the initial mass of the secondary was about
$M_{2i}\sim~ 0.63 M_{\sun}$. In other words, $0.29M_{\sun}$ have already been evaporated or transferred to the primary star. 
Considering $M_1=1.35 M_{\sun}$ and the accretion rate of about $10^{-8} M_{\sun}~yr^{-1}$, CAL 87 may explode as a type Ia supernova in $10^{7}~yr$.
The parameters of CAL 87 are presented in Table~\ref{parameters}.

Figure~\ref{fig1} presents the $P/\dot{P}$ \textit{versus} $P$ relation for WDMT and DIMT systems. The curve for WDMT systems at the beginning of wind-driven evolution was calculated from Eqs. (\ref{eq:um}), (\ref{eq:tres}) and (\ref{eq:cinco}) and from the values of $\beta_2$ and $q$ derived above for CAL 87. The parameter $P/\dot{P}$ is highly sensitive to the value of $\beta_2$, so the curve plotted is not extensive to other WDMT systems. Each WDMT evolutionary track was derived from Eq. (\ref{eq:um}) and Eq. (\ref{eq:quatro}). To construct the curves of the DIMT systems, we assumed a mass-radius relation for main sequence stars for the secondary, and the expression for the radius of the Roche lobe of the secondary from Paczynski (\cite{pacz}), and derived the orbital period through the determination of the separation between the binary components. The values of $P/\dot{P}$ were obtained assuming mass and angular momentum conservation in binary systems

\begin{equation}
\label{eq:seis}
P/\dot{P} = \frac{M_1 M_2}{3\dot{M}_{2}(M_2-M_1)}   
\end{equation}

\noindent and adopting the thermal timescale mass transfer for CBSS (Kahabka \& van den Heuvel \cite{kah}):

\begin{equation}
\label{eq:sete}
\dot{M}_{2} = \frac{-M_{2}^3}{3\times 10^{7}~M_2}.   
\end{equation}

Figure~\ref{fig1} illustrates the distinct behavior of the period change in the two regimes, WDMT and DIMT.

For a mass of $M_2 = 0.34 M_{\sun}$ for the secondary, the predicted radial velocity semi-amplitude is $K_1 = 68$ km s$^{-1}$. This is consistent with the amplitude of the radial velocity as determined by the absorption lines, but not consistent with the emission lines amplitude (Hutchings et al. \cite{hutchings}, Cowley et al. \cite{cow98}). Absorption lines also have the right phasing but large uncertainties. The emission lines have smaller amplitudes; \ion{He}{II}~4686~{\AA} has $K_1=30\pm{8}~\mathrm{km~s}^{-1}$ (Hutchings et al. \cite{hutchings}) and $40\pm{12}~\mathrm{km~s}^{-1}$ (Cowley et al. \cite{cow}), while \ion{O}{VI} presents $K_1=35 \pm{26}~\mathrm{km~s}^{-1}$ (Hutchings et al. \cite{hutchings}), but all are out of phase by about $\Delta\phi = 0.14$. As all the lines have different amplitudes and phasing, we conclude that they result from  a combination of absorption lines superposed on emission, distorting their profiles; non-orbital motions in the disc, and wind emission, may also  significantly affect the determination of the radial velocity amplitudes. Further studies are necessary to answer this question.

\begin{table}[!h]
\caption{Physical parameters and orbital elements of CAL 87.}
\label{parameters}
\begin{flushleft}
\begin{tabular}{lll}
\hline
\hline\noalign{\smallskip\smallskip}
            \noalign{\smallskip}
 Parameter   & Value & Reference \\
\hline
\noalign{\smallskip}
$M_1$  	& $1.35   M_{\sun}$   & Starrfield et al. \cite{starr} \\
$M_2$	& $0.34   M_{\sun}$   & this work\\
$q$	& 0.25        		& this work\\
$i$	& $78 \degr$  & Meyer-Hofmeister et al. \cite{meyer}\\
$P_{orb}$	& 10.6 hr  & Alcock et al. \cite{alco}\\
$M_{2i}$	& $0.63M_{\sun}$	& this work\\
$P_i$	& 5.7 hr	& this work\\
$\beta_2$	& 0.89	& this work\\
$P/\dot{P}$	&$+7.2(\pm1.3) \times 10^{6}$ years	& this work\\
$T_2	$	& $<1200$ K & this work\\
\noalign{\smallskip}
\hline
\end{tabular}
\end{flushleft}
\smallskip\noindent
\end{table}

\begin{table}[!h]
\caption{CBSS with measured $P/\dot{P}$.}
\label{measured}
\begin{flushleft}
\begin{tabular}{llll}
\hline
\hline\noalign{\smallskip\smallskip}
            \noalign{\smallskip}
 Star   &$P_{orb}$  & $P/\dot{P}$ & Reference\\
        & (hr) &        ($\times 10^6$ years) & \\
\hline
\noalign{\smallskip}
V617 Sgr	&  4.97	&  $+1.1\pm{0.2}$  & Steiner et al. \cite{stei06}\\
CAL 87		& 10.62	&  $+7.2\pm{1.3}$  & this work\\
V Sge		& 12.34	&  $-5.4\pm{0.2}$  & Patterson et al. \cite{patter},\\
		&	&		   & Szasz et al. \cite{szasz}\\
\noalign{\smallskip}
\hline
\end{tabular}
\end{flushleft}
\smallskip\noindent
\end{table}

As the system presents deep eclipses, one could argue that, at mid-eclipse, the spectrum of the secondary star could reveal its evolutionary status.
However, this may not be the case as its temperature is likely to be very low.  The expected temperature is
$T_2 < 1200~K$. This value is derived assuming that the luminosity is smaller than the value for the main sequence of this mass, and considering the radius derived from the current system parameters.

The inclination of the system was determined by Meyer-Hofmeister et al. (\cite{meyer}) assuming a secondary more massive than the primary. They found $i= 78 \degr$. In the current scenario the inclination is higher, being $i>75\degr$ in order for the white dwarf to be eclipsed, but more likely $i \simeq 79 \degr$ in order to produce the deep eclipse. The inclination cannot be much larger than this, otherwise we would not detect significant X-ray emission outside the eclipse because of strong attenuation by the accretion disc.

\begin{table}[!h]
\caption{CBSS with estimated mass ratios.}
\label{massratio}
\begin{flushleft}
\begin{tabular}{lllll}
\hline
\hline\noalign{\smallskip\smallskip}
            \noalign{\smallskip}
 Star &Type  &$P_{orb}$ (hr)     & $q=M_2/M_1$ & Refs.$^{\mathrm{a}}$\\
\hline
\noalign{\smallskip}
RX J0537.7-7034 & CBSS  & 3.48  & 0.5  & 1      \\
1E0035.4-7230   & CBSS  & 4.10  & 0.3  & 2      \\
V617 Sgr 	& V Sge & 4.97  & 0.4  & 3, 4   \\
WX Cen   	& V Sge & 9.98  & 0.35 & 5, 6   \\
CAL 87   	& CBSS  & 10.62 & 0.25 & 7,8      \\
V Sge    	& V Sge & 12.34 & 3.8  & 9,10   \\
QR And   	& CBSS  & 15.84 & 0.4  & 11, 12 \\
CAL 83   	& CBSS  & 24.98 & 0.4  & 13     \\
\noalign{\smallskip}
\hline
\end{tabular}
\end{flushleft}
\smallskip\noindent
\begin{list}{}{}
\item[$^{\mathrm{a}}$] References: 1- Greiner et al. \cite{greiner}; 2- van Teeseling et al. \cite{teesetal}; 3- Steiner et al. \cite{stei99}; 4- Steiner et al. \cite{stei06}; 5- Diaz \& Steiner \cite{diaz95}; 6- Oliveira \& Steiner \cite{oliv}; 7- Hutchings et al. \cite{hutchings}; 8- this work; 9- Herbig et al. \cite{herbig}; 10- Smak et al. \cite{smak}; 11- McGrath et al. \cite{mcgrath}; 12- Deufel et al. \cite{deufel}; 13-  Cowley et al. \cite{cow98}.
\end{list}
\end{table}

Are there other evolved WDMT systems among the long period CBSS, or is CAL 87 an exception? Cowley et al. (\cite{cow98}) suggested that \object{CAL 83} has a low-mass secondary, given its low amplitude radial velocity. Similarly, \object{WX Cen} shows indications that it has a low mass ratio (Diaz \& Steiner \cite{diaz95}; Oliveira \& Steiner \cite{oliv}). From the known systems that belong to the class of CBSS and V Sge stars, six are short period ($<6$ hr) systems (\object{GQ Mus}, \object{T Pyx}, \object{RX J0439}, \object{RX J0537}, \object{IE 0035}, and \object{V617 Sgr}) and are therefore WDMT systems. Among the long period objects, four have indications of being WDMT systems (\object{WX Cen}, \object{CAL 87}, \object{QR And} and \object{CAL 83}, see Table~\ref{massratio}), and one has no indication to date of its mass (\object{RX J0513}), while \object{MR Vel} probably has a subgiant secondary (Schmidtke et al. \cite{schmidtke3}).

The only clear case for a DIMT to date is \object{V Sge} (Herbig et al. \cite{herbig}; Patterson et al. \cite{patter}). This supports the argument that wind-driven mass transfer (with ten objects) may be the rule and dynamical instability mass transfer (with one object), the exception. Another implication is that the CBSS are an old population, instead of a population of intermediate age as previously believed.

\begin{acknowledgements}
A. S. Oliveira acknowledges FAPESP (grant 03/12618-7) for financial support. We would like to thank Dr. Albert Bruch for his careful reading of the manuscript.
\end{acknowledgements}

\end{document}